\documentclass[twocolumn]{aastex631}

\usepackage{siunitx}
\usepackage{natbib}
\usepackage{bm}
\usepackage{mathtools}
\usepackage{multirow}
\usepackage{CJK}         % Chinese name
\usepackage{hyperref}

\begin{document}
\begin{CJK*}{UTF8}{gbsn}

\title{Measuring the Vertical Structure of Active Galactic Nuclei Disks with Transformer Models and the Vera C. Rubin Observatory}

%\linenumbers

\author[0000-0002-1174-2873]{Amy Secunda}\thanks{E-mail: asecunda@flatironinstitute.org}
\affil{Center for Computational Astrophysics, Flatiron Institute, 162 Fifth Avenue, New York, NY 10010, USA}

\author[0000-0001-5039-1685]{Sebastian Wagner-Carena}
\affil{Center for Computational Astrophysics, Flatiron Institute, 162 Fifth Avenue, New York, NY 10010, USA}

\author[0000-0003-1899-9791]{Helen Qu}
\affil{Center for Computational Astrophysics, Flatiron Institute, 162 Fifth Avenue, New York, NY 10010, USA}

\author[0000-0002-1068-160X]{Shirley Ho}
\affil{Center for Computational Astrophysics, Flatiron Institute, 162 Fifth Avenue, New York, NY 10010, USA}
\affil{Department of Astrophysical Sciences, Princeton University, 4 Ivy Lane, Princeton, NJ 08544, USA}
\affil{Department of Physics \& Center for Data Science, New York University, 726 Broadway, New York, NY 10003, USA}

\begin{abstract}
Reverberation mapping is one of the main techniques used to study active galactic nuclei (AGN) accretion disks. Traditional continuum reverberation mapping uses short lags between variability in different wavelength AGN light curves on the light crossing timescale of the disk to measure the radial structure of the disk. The harder-to-detect long negative lag measures lags on the longer inflow timescale, opening up a new window to mapping out the vertical structure of AGN disks. The Vera Rubin Observatory, with its 6 wavebands, long baseline, and high cadence, will revolutionize our ability to detect short and long lags. However, many challenges remain to detect these long lags, such as seasonal gaps in Rubin light curves, the weak signal strength of the long lag relative to the short lag, and the enormous influx of data for millions of AGN from Rubin. Machine learning techniques have the potential to solve many of these issues, but have yet to be applied to the long negative lag problem. We develop and train a transformer-based machine learning model to detect long and short lags in mock Rubin AGN light curves. Our model identifies whether a light curve in our test set has a long negative lag with 96\% recall and $0.04\%$ contamination, and is 98\% accurate at predicting the true long lag. This accuracy is an enormous improvement over two baseline methods we test on the same mock light curves, the interpolated cross correlation function and {\sc javelin}, which are only 54\% and 21\% accurate, respectively.
\end{abstract}

\section{Introduction}
\label{sec:intro}

Active Galactic Nuclei (AGN) accretion disks fuel radiation and outflows that can impact galaxy evolution on large scales \citep{Fabian2012,Kormendy2013}. AGN disks also provide an important laboratory for studying accretion physics. Unfortunately, due to their great distance, it is not possible to spatially resolve the vast majority of AGN accretion disks. Instead, most of what we observe about AGN disk structure comes from a technique known as disk continuum reverberation mapping, which was first adapted by \cite{Collier:1999} from a similar idea for mapping the AGN broad line region used by \cite{Blandford:1982}. 

Continuum reverberation mapping is often interpreted in the context of the lamp-post model. In the lamp-post model, time-varying X-ray radiation is emitted by the corona and moves radially outward on the light crossing timescale to the UV- and optical-emitting regions of the disk where it gets absorbed, reprocessed, and re-emitted at different wavelengths based on the local black body temperature of the disk. The reprocessing of X-ray variability causes variability in longer wavelength light curves to lag variability in shorter wavelength light curves. Measuring these lags, which we will call short lags here, allows us to measure the radial temperature profile and radial extent of an AGN disk \citep[e.g.,][]{Sergeev:2005,Cackett:2007,Cackett:2018,derosa2015,Edelson:2015,Edelson2017,Edelson:2019,Jiang:2017,Fausnaugh:2016,Starkey:2017,Homayouni:2022}.

%in the inner hotter regions of an AGN disk moves outwards to the outer cooler disk on the light-crossing timescale. As they move outward, these X-ray photons are absorbed, reprocessed, and re-emitted  at different wavelengths based on the local black body temperature of the disk. This reprocessing of X-ray variability leads variability in longer wavelength light curves to lag variability in shorter wavelength light curves. Finding these lags allows us to measure the radial temperature profile and radial extent of an AGN disk \citep[e.g.,][]{Sergeev:2005,Cackett:2007,Cackett:2018,derosa2015,Edelson:2015,Edelson2017,Edelson:2019,Jiang:2017,Fausnaugh:2016,Starkey:2017,Homayouni:2022}.

The lamp-post model assumes that reprocessed X-ray variability is the main driver of variability in UV and optical light curves. However, there is growing evidence of intrinsic variability not driven by X-ray variability in UV-optical AGN light curves \citep[e.g.,][]{Arevalo:2009,Neustadt:2022,F92020,Beard:2025} and a number of examples of AGN with only weak to moderate correlations between X-ray and UV-optical light curves \citep[e.g.,][]{Schimoia:2015,Buisson:2018,Edelson:2019,Cackett:2023,Kara:2023}. Recent radiation magnetohydrodynamic simulations suggest that fluctuations in the AGN disk driven by the magnetorotational instability \citep[MRI,][]{BalbusHawley1991} and convection can produce intrinsic variability in AGN light curves \citep{JiangBlaes2020,Secunda:2024,Secunda:2025,Kaul:2025}. Identifying this intrinsic variability in AGN light curves could allow us to probe the internal physics of AGN disks.

Also tantalizing is the prospect of being able to measure long negative lags between intrinsic variability in different waveband light curves coming from the inward propagation of fluctuations in AGN disks on the viscous or inflow timescale of the disk. In the standard \cite{ShakuraSunyaev1973} thin disk model, the inflow timescale can be parameterized as
\begin{equation}
    \label{eq:inflow_param}
    \tau_{\rm{inflow}} = 260 \frac{M}{10^8 M_{\odot}} \frac{0.1}{\alpha} \left(\frac{h}{0.01r}\right)^2 \left(\frac{r}{30r_{\rm{g}}}\right)^{3/2}\rm{~years},
\end{equation}
where $M$ is the supermassive black hole mass, $\alpha$ is the \cite{ShakuraSunyaev1973} stress parameter, $h/r$ is the aspect ratio, where $h$ is the scale height of the disk and $r$ is the disk radius, and $r_g$ is gravitational radii. In the thin disk model, $h/r=0.01$ leading to an inflow timescale on the order of 100 years, or far too long to detect in current AGN light curves. However, there is growing evidence from observations \citep[e.g.,][]{Morgan:2010,Jiang:2017,Guo:2022} and simulations \citep[e.g.,][]{Gaburov:2012,Jiang2016,Jiang:2019,JiangBlaes2020,Hopkins:2023} that the standard thin disk model does not fully describe an AGN disk \citep[see also,][]{Koratkar:1999,Antonucci:1988}. Because the inflow timescale depends on the height of the disk, long negative lags can be used to measure the height of AGN disks and serve as a fundamental test of the standard thin disk model.

%that the standard thin disk model does not fully describe an AGN disk. For example, both micro-lensing and reverberation mapping campaigns suggest that the radial extent of the disk is larger than predicted by the standard thin disk model \citep[e.g.,][]{Morgan:2010,Jiang:2017,Guo:2022}. In addition, recent simulations of AGN disks suggest that pressure support from magnetic fields and radiation can lead to puffed up or slim instead of thin disks \citep[e.g.,][]{Gaburov:2012,Jiang2016,Jiang:2019,JiangBlaes2020,Hopkins:2023}.  \citep[see also,][]{Koratkar:1999,Antonucci:1988}.

There is already some evidence for long negative lags in AGN light curves. \cite{Neustadt:2022} and others measured propagating temperature fluctuations in AGN photometry and spectroscopy \citep[see also,][]{Neustadt:2023,Stone:2023}. \cite{Yao:2022} directly measured a long negative lag ranging from around $-10$ to $-70$~days between several different waveband light curves for the AGN Fairall 9. This long negative lag suggests an aspect ratio of $h/r\approx0.2$ \citep{Secunda:2023}, significantly thicker than is typically assumed in the standard thin disk model. 

Observing more long negative lags is crucial to develop better models for these lags and understand how they can be used to study AGN disk structure. The Vera Rubin Observatory Legacy Survey of Space and Time, which has 6 wavebands, a planned 10-year baseline, and will contain over 100 million AGN \citep{Ivezi2019}, is the ideal instrument for detecting long negative lags. Unfortunately, \cite{Secunda:2023} tested several traditional reverberation mapping methods on mock Rubin light curves, and found that the roughly consistent $\sim 6$ month seasonal gaps in these light curves will lead to problems with aliasing and false signal detections on timescales comparable to the long negative lag. In addition, \cite{Secunda:2023} found that the confounding signal from the short lag, which should in most cases be a stronger signal than the long lag signal in order to be consistent with observations, prevents traditional methods from accurately detecting the long negative lag consistently.

%An in depth discussion of the long negative lag and the prospects for detecting them in Rubin light curves using traditional lag detection methods is provided in \cite{Secunda:2023}.

In this paper, we develop a machine learning method with a transformer-based architecture \citep{Vaswani2017} to predict the posterior distribution over the short and long lag. We train our transformer on a test set of mock Rubin AGN light curves in preparation for applying our model to real Rubin AGN light curves as they become available in the coming years. While training, our model implicitly learns to marginalize over the uncertainty introduced by missing data (gaps) and the confounding signal (short lag). Although transformers have not yet been widely applied to the analysis of AGN light curves, previous studies have used other machine learning methods to analyze real and mock AGN light curves to model AGN variability \citep[e.g.,][]{Tachibana:2020,Sanchez-Saez:2021,Hajdinjak:2022,Sheng:2022,Danilov:2022} and predict supermassive black hole and disk properties using AGN light curves \citep[e.g.,][]{Park:2021,Li:2024,Fagin:024,Best:2024b,Jimnez-Vicente:2025}. However, none of these methods model the long negative lag. In fact, the mock light curves they are trained on do not include any intrinsic variability, guaranteeing that a long negative lag will not be present.

This paper is organized as follows. In Section \ref{sec:methods} we describe our method for simulating these mock light curves (Section \ref{sec:methods:sims}), the architecture of our transformer (Section \ref{sec:methods:transformer}), and two baseline long lag detection methods we use for comparison (Section \ref{sec:methods:other}). We show our model's predictions for a test set of mock Rubin light curves in Section \ref{sec:results:transformer} and compare it to our baseline methods in Section \ref{sec:results:compare}. Finally, we summarize our findings in Section \ref{sec:conclude}.

% density estimator, a posterior estimator, a neural posterior estimator, or posterior surrogate model

\section{Methods}
\label{sec:methods}

\subsection{Simulating Mock Light Curves}
\label{sec:methods:sims}

We simulate the mock Rubin light curves that we use to train our transformer model using similar methods to \cite{Secunda:2023}. We highlight the most important details in this section. First, we generate a driving light curve with {\sc EzTao} \citep{Yu:2022}. {\sc EzTao} is a Python toolkit that uses {\sc celerite}, a fast Gaussian process regression library, to generate different representations of AGN variability. Here we model our light curves as a damped random walk (DRW). The DRW is a stochastic process with a frequency, $\nu$, dependent power spectral density that goes from  $\propto \nu^{-2}$ at high frequencies to white noise at frequencies below a characteristic damping timescale, $\tau_{\rm damp}$.  The amplitude of the DRW is set by the structure function at infinity, SF$_{\infty}$. The DRW is a good representation of AGN variability on timescales of days to years \citep[e.g.,][]{Kelly:2009,Kozlowski2010,MacLeod:2010,Zu2013,Burke2021}. 

We add a long and short lag to our light curves by reprocessing the driving light curve with two Gaussian response functions of the form,
\begin{equation}
\label{eq:gauss}
    \psi(t) = \frac{1}{S\sqrt{2\pi}}\exp{\frac{-(t-\tau_{\text{short/long}})^2}{2S^2}},
\end{equation}
where $S$ is the width of the Gaussian. For a negative lag we replace $t$ with $-t$. We randomly draw our true short lag from a uniform distribution, $\tau_{\rm short} \in [0.05,10]$ days and our true long lag from a uniform distribution, $\tau_{\rm long} \in [-3650,-10]$ days. If $\tau_{\rm long}<-1825$, we set $\tau_{\rm long}=0$ by only reprocessing our driving light curve with a single Gaussian for the short lag. 

Since the duration of long lags is not well constrained, we set our potential range as broad as possible, from $-10$ days to $-1825$ days, or about half the length of the ten year mock Rubin light curve. We also set roughly half of our light curves to $\tau_{\rm long}=0$, because we do not expect all AGN light curves to have a long negative lag and we want to train our model to only predict long negative lags if they are actually present. We also set a ratio, $r$, of the amplitude of the long negative lag to the amplitude of the short lag, by renormalizing the long lag response function. We expect that in most, if not all, cases the short lag will be a stronger signal than the long lag. Therefore, we draw $r$ from a log-uniform distribution $r \in [10^{-3},1]$. 

Once we have our driving and reprocessed light curves the final step is to mock observe these light curves with the Rubin cadence and errors using the Rubin Operations Simulator ({\sc OpSim}) 10 year baseline version 4.3.1 \citep{Reuter:2016}. Rubin will have two main survey types, the wide fast deep survey and the deep drilling fields. The former will have larger sky coverage, while the latter will have greater depth and higher cadence. For now, we focus on the deep drilling fields to utilize the higher cadence, which should be helpful in accurately recovering long negative lags. We mock observe the driving and reprocessed light curves using the Rubin \emph{g}- and \emph{z}-band cadences, respectively, for a randomly selected deep drilling field sky position. We choose these bands because they provide the widest range in wavelength, except for the \emph{u}- and \emph{y}-bands, which have fewer total observations. Because the scaling of the long negative lag across wavelength is poorly understood, we use only two wavebands at a time. We also add the band-dependent dust extinction and Gaussian noise to our mock Rubin light curves by calculating the signal-to-noise ratio (S/N) for each observation using the sky brightness and instrument noise from {\sc OpSim}. 

We show the range of parameters we use to generate our mock light curves in Table \ref{tab:params} of Appendix \ref{app:methods}. The broad range of parameters reflects the current uncertainty in the long negative lag. Although several of these parameters may be correlated with each other, because the exact correlations of many parameters are uncertain, we conservatively do not allow our model to learn these correlations to improve its constraining power. We train our model on $\num{2e6}$ light curves in order to span all possible parameter space.

\subsection{Transformer Model}
\label{sec:methods:transformer}

%that has become standard in natural language processing, because it is excellent at learning the conditional distribution of data. Transformers have already been applied to several problems in astronomical time-series data, such as light curve classification for Rubin \citep[e.g.,][]{Cabrera-Vives:2024,Gupta:2025,Moreno:2025}, stellar light curve analysis \citep[e.g.,][]{Pan:2022, Salinas:2023,Leung:2024,Chiong:2025,Cadiz:2025}, and supernovae classification \citep[e.g.,][]{Pimentel:2023,qu2024connect}, but have not been specifically applied to AGN light curve analysis yet. The self-attention mechanism allows transformers to learn the long-range dependencies of our light curves without sequential processing, making it fast to train on a large data set of mock Rubin light curves and highly accurate. 9, 

The goal of our transformer is to learn the posterior over the short and long lag distribution in order to accurately predict these lags. We apply a transformer model to this problem because transformers have been shown to perform well on astronomical time series data, such as light curve classification for Rubin \citep[e.g.,][]{Cabrera-Vives:2024,Salinas:2023,qu2024connect}. The input to our model is the time and brightness of each light curve, which we pre-process by (1) normalizing our light curves to vary between 0 and 1, (2) combining observations down to the nearest daily interval, and (3) setting unobserved time steps to $-2$. (2) makes the problem more tractable and intranight observations should not be particularly helpful in detecting long negative lags that are on timescales of $10-2000$~days. (3) allows us to aggregate the information encoded in our light curve using global average pooling.

The architecture of our optimized model is,
\begin{itemize}
    \item An embedding layer with time-based positional encoding.
    \item 8 layers of transformer blocks with 256 dimensions and 16 heads each.
    \item A global average pooling.
    \item A dense layer.
\end{itemize}
Each transformer block consists of a normalization layer, a multi-head dot product attention, a second normalization layer, and a multi-layer perceptron. 

We parameterize our posterior as a mixture of two Gaussians, $\mathcal{N}\left(y_{\phi=0,1} | \mu_{\phi=0,1}(x),\Sigma_{\phi=0,1}(x) \right)$. $\phi=1$ represents the case where there is no long lag, $y_{\phi=1}$ and $\mu_{\phi=1}(x)$ are the true and predicted short lag only, respectively, and $\Sigma_{\phi=1}$ is the precision matrix for the short lag only. $\phi = 0$ is the case when there is a long lag, $y_{\phi=0}$ is the true and $\mu_{\phi=0}(x)$ is the predicted long and short lag, and $\Sigma_{\phi=1}$ is the precision matrix for the long and short lag.

The log loss,
\begin{equation}
\begin{split}
\log L &= (1-\phi)\log(1 - p) + \phi \log(p) \\
&\quad +  \log N\left(y_\phi \mid \mu_{\phi}(x), \Sigma_{\phi}(x)\right)
\end{split}
\end{equation}
comes from minimizing the KL divergence with the true distribution. Here $p$ weights the two Gaussians and is the probability of there being a long lag. This loss function allows us to learn the probability of a long lag and the distribution of the short and long lag conditioned on the presence of the long lag. We use the Adam optimizer with a cosine decay and a batch size of $256$.

We perform a Bayesian optimization for the hyper-parameter ranges given in Table \ref{tab:hyper} in Appendix \ref{app:methods} with 75 trials on wandb using the Tree-structured Parzen Estimator algorithm to pick the next point \citep{Bergstra:2011}. We train our model on two million mock driving and reprocessed Rubin light curves. Our model posterior can be factorized into three components for each light curve, the probability of there being a long negative lag, a prediction for the long negative lag, and a prediction for the short lag. We determine the optimal model by minimizing the loss in a validation sample of $2 \times 10^4$ mock light curves. Our optimized model took 13~hours and 11~minutes to train on two million light curves on a single NVIDIA A100-SXM4-40GB GPU.

\begin{figure}
    \centering
    \includegraphics[width=\linewidth]{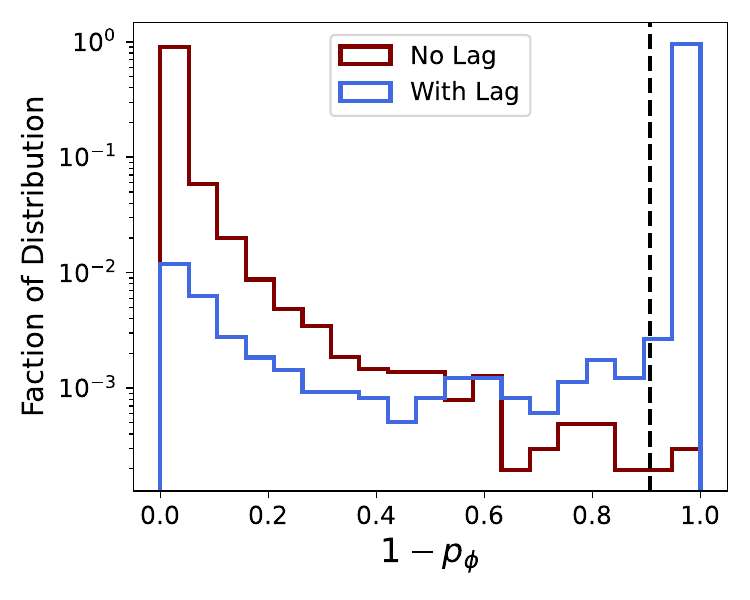}
    \caption{The distribution of the predicted likelihood of a long negative lag ($1-p_\phi$) in our test set mock light curves without (in brown) and with (in blue) long negative lags. The dashed black line shows our cut off, $(1-p_\phi)>0.906$, for filtering out light curves without long lags. This cut-off gives us a sample with $0.04\%$ contamination and 96\% completeness.}
    \label{fig:p_phi}
\end{figure}

\begin{figure*}
    \centering
    \includegraphics[width=0.945\linewidth]{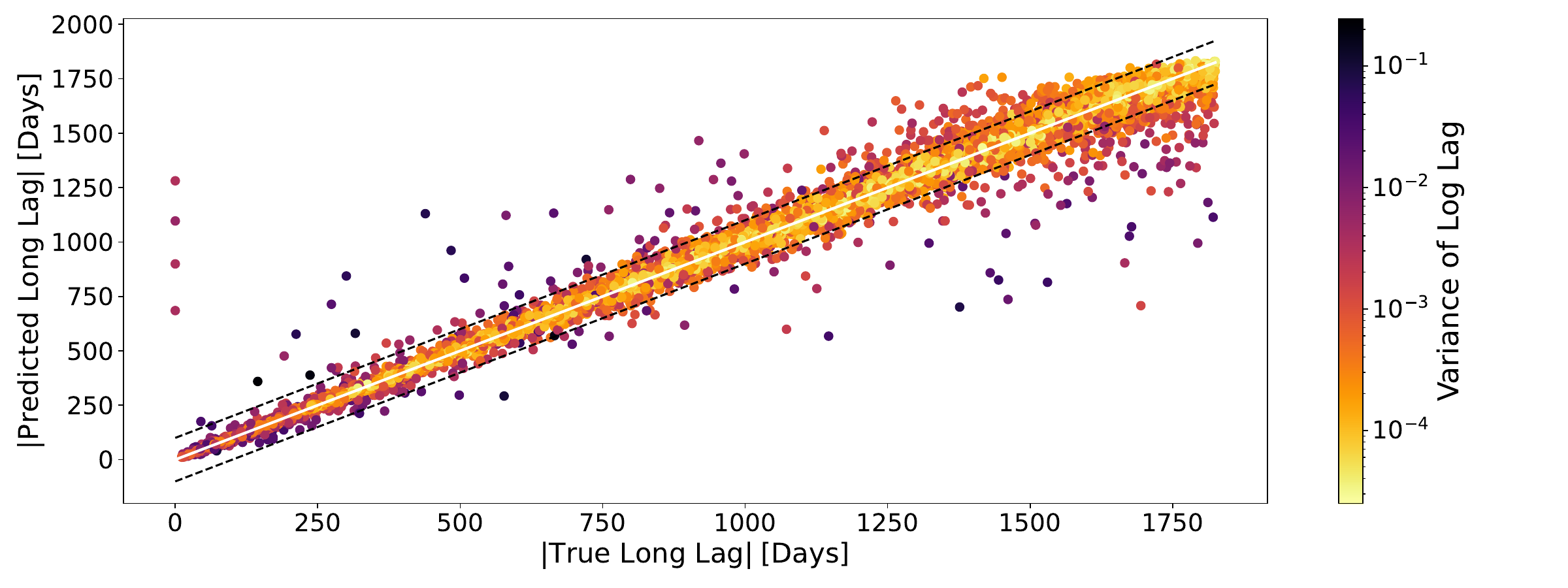} \\
    \includegraphics[width=0.945\linewidth]{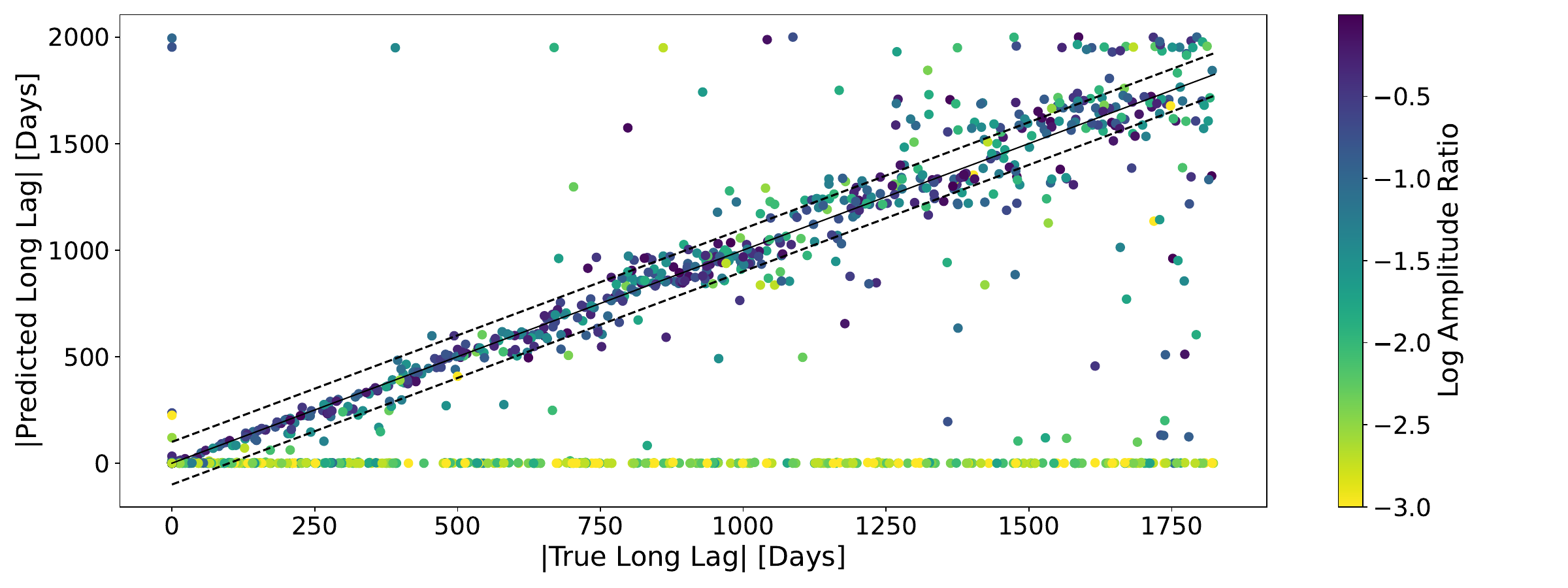} \\
    \includegraphics[width=0.945\linewidth]{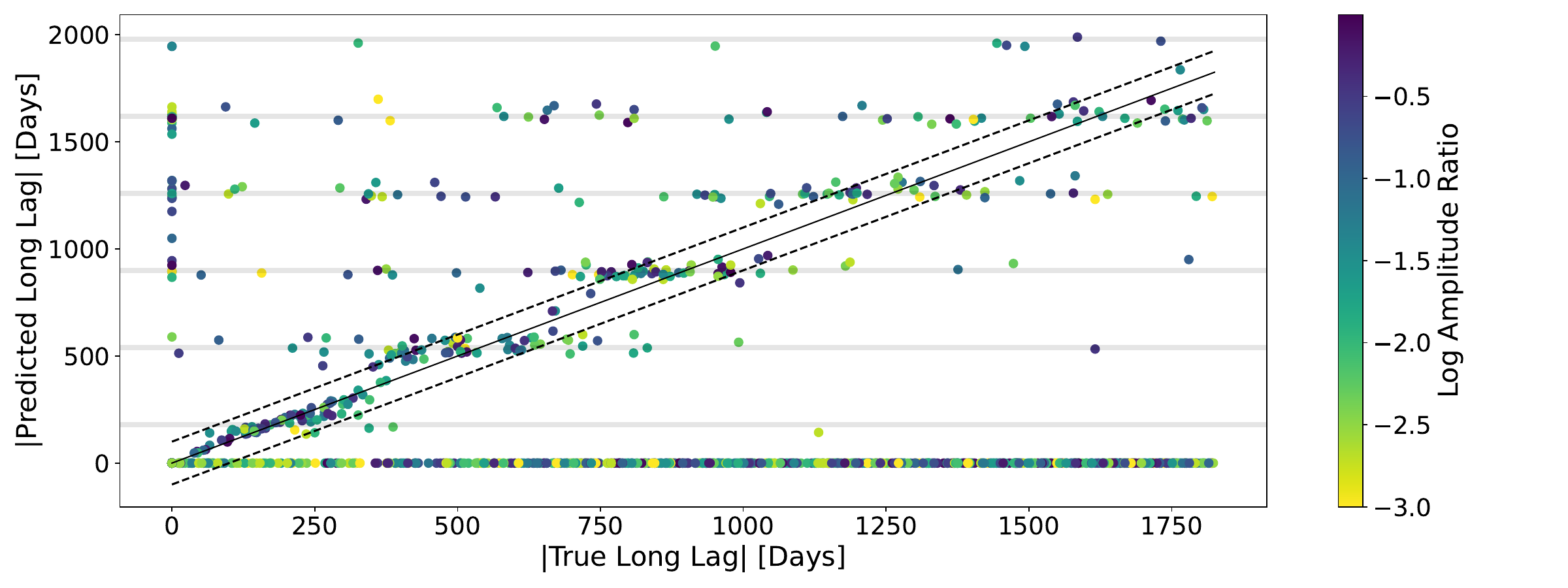}
    \caption{The top panel shows the duration of the true long lag versus the duration of the long lag predicted by our transformer model for our test set of $\num{2e4}$ mock light curves. The middle and bottom panel show the duration of the true long lag versus the duration of the long lag predicted by the ICCF and {\sc javelin}, respectively, for a subset of 2000 light curves from our test set. The solid lines are the one-to-one lines for a perfect prediction, the dashed lines show $\pm 100$ days. In the top panel the color bar shows the variance of the posterior for the logarithm of the long lag. In the middle and bottom panels the color bar shows the logarithm of the ratio of the amplitude of the long lag to the amplitude of the short lag. The horizontal gray lines in the bottom panel show odd integer multiples of 180~days which is roughly the seasonal gap in observations. 98\% of long lags predicted by our model are within 20\% of the true long lag. On the other hand, {\sc javelin} experiences severe aliasing problems and does not reliably predict the long lag, while the ICCF fails to predict any long lag for light curves with low amplitude ratios.}
    \label{fig:main}
\end{figure*}

\subsection{Other Lag Detection Methods}
\label{sec:methods:other}

In Section \ref{sec:results:compare} we compare the accuracy of our transformer to two standard methods used to detect long negative lags in AGN light curves, the interpolated cross-correlation function \citep[ICCF,][]{Gaskell1987} and {\sc javelin} \citep{Zu2011}. The ICCF linearly interpolates two light curves to an even cadence and then finds the correlation coefficient between them as a function of the delay time, $\tau$, 
\begin{equation}\label{eq:coeff}
    CC_{\rm coeff}(\tau) = \frac{\sum g(t)z(t+\tau)}{\sqrt{\sum g(t)^2 z(t+\tau)^2}},
\end{equation}
where $g(t)$ and $z(t)$ are the driving and reprocessed light curves, respectively. Here we consider the positive delay time where $CC_{\rm coeff}(\tau)$ peaks to be the short lag and the negative delay time where $CC_{\rm coeff}(\tau)$ peaks to be the long negative lag.

The second method is {\sc javelin} which uses a Monte Carlo Markov Chain (MCMC) to simultaneously derive the
posterior distributions for the light curves' DRW parameters and the time lag between the driving and reprocessed light curves. To determine the long negative lag we take the median of the negative {\sc javelin} distribution as in \cite{Secunda:2023}, but set a threshold that at least 60\% of the total {\sc javelin} distribution is less than zero. Distributions below that threshold are assigned a long negative lag of zero. A lower threshold leads to higher completion, but also significant contamination. A higher threshold leads to very few lag detections. For the short lag, we use the median of the positive {\sc javelin} distribution.

\section{Results}
\label{sec:results}

\subsection{Transformer Model Performance}
\label{sec:results:transformer}

\begin{figure}
    \centering
    \includegraphics[width=\columnwidth]{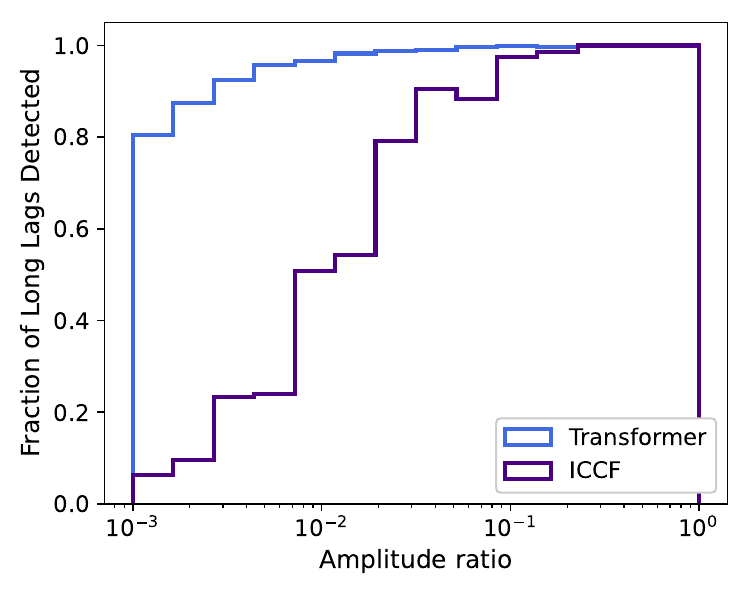}
    \caption{The fraction of light curves with long negative lags detected in each amplitude ratio bin for our transformer model (in blue) and the ICCF method (in purple). The transformer model detects long lags for all amplitude ratios only missing 19\% of long lags for the lowest amplitude ratio bin, while the ICCF fails to detect 87\% of long lags for light curves with amplitude ratios, $r<0.01$.}
    \label{fig:missing}
\end{figure}

We assess the accuracy of our transformer model on a test set of $\num{2e4}$ mock light curves in Figure \ref{fig:p_phi} and the top panel of Figure \ref{fig:main}. Figure \ref{fig:p_phi} shows the distribution of the predicted likelihood ($1-p_\phi$) of a light curve having a long negative lag for mock light curves with (in blue) and without (in brown) long negative lags. Our model predicts $(1-p_\phi)<0.1$ for 95\% of our test set light curves without a long negative lag and $(1-p_\phi)>0.99$ for 95\% of test set light curves with a long negative lag. 

We use our validation set of $\num{2e4}$ mock light curves to determine a lower limit for $(1-p_\phi)$ that filters out light curves without long negative lags. For our validation set, choosing $(1-p_\phi)>0.906$ gives us a sample with zero contamination and 96\% completeness. In other words, none of our validation set light curves without long lags have $(1-p_\phi)>0.906$ and 96\% of our validation set light curves with long lags have $(1-p_\phi)>0.906$. If we use this limit to select out light curves with long lags in our test set, we find 96\% completion and only 0.04\% contamination.

The blue distribution in Figure \ref{fig:missing} shows the fraction of test light curves with long lags that our model predicts $(1-p_\phi)>0.906$ for. Unsurprisingly, the highest failure rate occurs at the lowest amplitude ratios, for which the long lag signal is the weakest. However, even in the lowest amplitude ratio bin, 81\% of long lags are detected and only 13\% of light curves with $r<0.01$ are missed by our model.

The top panel of Figure \ref{fig:main} shows the duration of the true long lag versus the duration of the long lag predicted by our model for mock light curves in our test set with $(1-p_\phi)>0.906$. The points are colored by the variance of the posterior for the logarithm of the long lag for that light curve. 98\% of long negative lags predicted by our model are accurate to within 20\% of the true lag and 92\% are accurate to within 10\% of the true lag. In addition, 94\% of predicted lags are within $\pm100$~days of the true lag. Because the width of the response function we reprocess our lags with (see Section \ref{sec:methods:sims} and Table \ref{tab:params}) scales with the length of the long negative lag, our model predictions are more accurate for shorter long negative lags. There is also a clear color gradient as you move outward from the one-to-one line shown in white, showing that the most accurate lag predictions have low uncertainties and the least accurate lag predictions tend to have higher uncertainties. 

\begin{figure}
    \centering
    \includegraphics[width=\columnwidth]{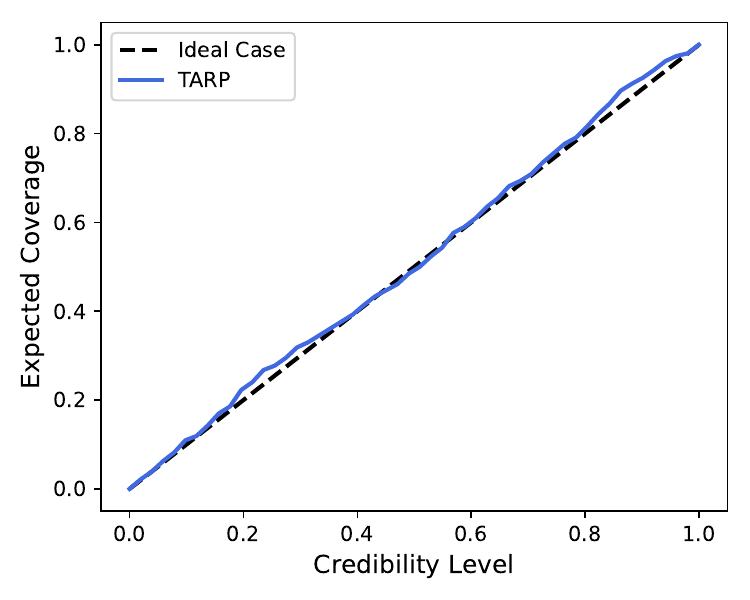}
    \caption{Our TARP calculation (in blue) of the expected coverage probability versus the credibility level for our model posterior. The dashed black line indicates perfect calibration. Our model does very well on this calibration test, with slight under-confidence in a few quantiles.}
    \label{fig:tarp}
\end{figure}

While we visually find that the uncertainty in our model corresponds well to the accuracy of the prediction, we quantitatively assess the calibration of our posterior estimator by performing a Tests of Accuracy with Random Points (TARP) coverage test \citep{Lemos:2023}. We show the expected coverage probability versus the credibility level for this calculation in Figure \ref{fig:tarp}. Our TARP calculation shows that our model posterior is well calibrated, which means that its uncertainty estimates match the empirical error distribution. Note that this TARP calculation includes predictions for the short lag, although we defer discussion of the accuracy of our model at predicting short lags to Appendix \ref{app:short}. Overall, our transformer model is highly accurate at predicting whether there is a long lag, predicting the duration of the long lag, and estimating the uncertainty of the long lag. 

%This high accuracy will allow us to confidently apply our model to real data, and is also far superior to the accuracy of other pre-existing methods used to find long negative lags, which we will now show in the following section.

\subsection{Comparison to Other Methods}
\label{sec:results:compare}

In this section, we compare the accuracy of our transformer model to two more traditional lag recovery methods described in Section \ref{sec:methods:other}, the ICCF \citep{Gaskell1987} and {\sc javelin} \citep{Zu2011}. The middle panel of Figure \ref{fig:main} shows the duration of the true long lag versus the duration of long lag predicted by the ICCF for 2000 randomly selected mock light curves from the same test set we use to examine the performance of our transformer model in the previous section. A nice feature of the ICCF results is that the ICCF prediction is only longer than 20~days for 0.6\% of mock light curves with no long lag. Therefore, we can use the ICCF to weed out false long lag detections. However, the ICCF fails to detect a long lag for about a third of light curves with a long lag and overall is accurate to within 20\% of the long lag for only 54\% of light curves that have a long negative lag. 

We compare the fraction of light curves with long lags detected as a function of the amplitude ratio for our transformer model and the ICCF in Figure \ref{fig:missing}. The ICCF fails to detect long negative lags in 87\% of light curves where the ratio of the amplitude of the long lag to the amplitude of the short lag is $r\lesssim0.01$, while our transformer model only misses 13\% of light curves with $r\lesssim0.01$. Although the ICCF fails to detect a third of long negative lags, when we only consider light curves that the ICCF does detect long lags for, it is accurate to within 20\% of the true long negative lag 83\% of the time. 

On the other hand, the bottom panel of Figure \ref{fig:main} shows that the {\sc javelin} results heavily suffer from aliasing, over-predicting lags at odd integer multiples of the seasonal gaps, which we show as the horizontal gray lines in Figure \ref{fig:main}. As a result, {\sc javelin} is only able to predict accurate lags if they fall around these integer multiples or are on timescales less than a year. Overall, {\sc javelin} is only accurate to within 20\% of the true long lag for 21\% of light curves with a long negative lag and only finds a long lag for 34\% of mock light curves that have a long lag. In addition, 4.0\% of long lags are falsely detected by {\sc javelin} for light curves that have no long lags. Although 4\% is still small, it is an order of magnitude larger than the number of false detections with the ICCF and two orders of magnitude larger than the number of false detections with our transformer model. Finally, unlike with the ICCF, when we only consider light curves that {\sc javelin} does detect long negative lags for, the accuracy only increases to 54\%.

These results suggest that {\sc javelin} is not a very useful method for predicting long negative lags in Rubin AGN light curves. On the other hand, because it is reasonably accurate when able to detect lags, the ICCF could be used to corroborate long lags with $r>0.01$ predicted by our transformer model. Nevertheless, our transformer is significantly more accurate overall and is actually able to detect lags with low amplitude ratios. Our transformer is also dramatically faster. Once trained, our transformer can run on 1000 light curves on a single NVIDIA A100-SXM4-40GB GPU in 0.9 seconds, whereas it takes the ICCF and {\sc javelin} 12 hours and 5 days, respectively, to run on 1000 light curves on a single CPU.

\section{Discussion and Summary}
\label{sec:conclude}

We develop and train a transformer model to detect long negative lags up to a duration of 1825~days in mock Vera Rubin Observatory AGN light curves. Our model is able to accurately predict the presence of a long negative lag for 96\% of mock light curves in our test set that have long negative lags, with only 0.04\% contamination from light curves without long negative lags. In addition, our model accurately predicts a long negative lag to within 20\% of the true lag for 98\% of our test mock light curves. These results demonstrate that our model is able to handle the challenges of detecting long negative lags in Rubin light curves, including the uneven Rubin cadence and the weakness of the long lag signal relative to the short lag signal.

On the other hand, the ICCF and {\sc javelin} do not handle these challenges well. These traditional methods are only 54\% (ICCF) and 21\% ({\sc javelin}) accurate. The ICCF suffers from being unable to detect long negative lags if the signal from the short lag is significantly stronger than the signal from the long lag. {\sc javelin} suffers from aliasing issues that have been reported by others searching for short lags in real data sets \citep[e.g.][]{Grier:2017,Grier:2019,Shen:2024,McDougall:2025,Homayouni:2025}, but appear to be even worse when searching for long lags. In addition to being more accurate, our transformer model is also over $10^4 \times$ faster than the ICCF and $10^5 \times$ faster than {\sc javelin} which uses an MCMC. This speed up will be important for handling the enormous influx of AGN light curves from Rubin.

We caution that we only test our transformer model on mock light curves, generated the same way as our training and validation set. Despite adding Rubin-like cadences and errors, real Rubin light curves will likely present new challenges for our transformer model. We also do not account for anticipated correlations between several of the parameters of our mock light curves, such as the damping timescale and the duration of the long and short lags, which should all depend on the mass of the supermassive black hole. We have not discovered enough long negative lags to have a robust model for how these lags correlate with other parameters. We therefore conservatively prevent our model from learning these correlations to improve its constraining power. Instead, we train our transformer model on $2\times 10^6$ light curves in order to span all possible parameter space. 

While we await multi-year data from Rubin, future work should test our transformer model on light curves from the Zwicky Transient Facility, an optical survey telescope with up to 6~years of observations in 2--3 wavebands. As we observe more long negative lags, we will be able to develop a better constrained empirical model, which will allow us to refine our current long lag detection methods. Future transformers could also predict the posterior distribution for other AGN disk and light curve parameters. In particular, predicting the amplitude ratio, $r$, would be useful in modeling AGN variability and improving non-machine learning methods for detecting long and short lags.

Detecting long negative lags will allow us to track the propagation of fluctuations in AGN disks, providing novel information on the internal physics of the disk as well as the vertical structure of the disk. This information will allow us to test different models for AGN disks, from the \cite{ShakuraSunyaev1973} standard thin disk model to the magnetically-elevated AGN disks presented in \cite{Hopkins:2023}. Developing fast and accurate techniques for observing long negative lags, like the transformer model in this paper, is crucial to prepare for the numerous long baseline, high cadence, multi-waveband light curves coming from Rubin.

\begin{acknowledgements}
    The authors would like to thank Fran\c{c}ois Lanusse for his help with the transformer model. The Center for Computational Astrophysics at the Flatiron Institute is supported by the Simons Foundation. 
\end{acknowledgements}

\appendix

\begin{figure*}[b!]
    \centering
    \includegraphics[width=0.49\linewidth]{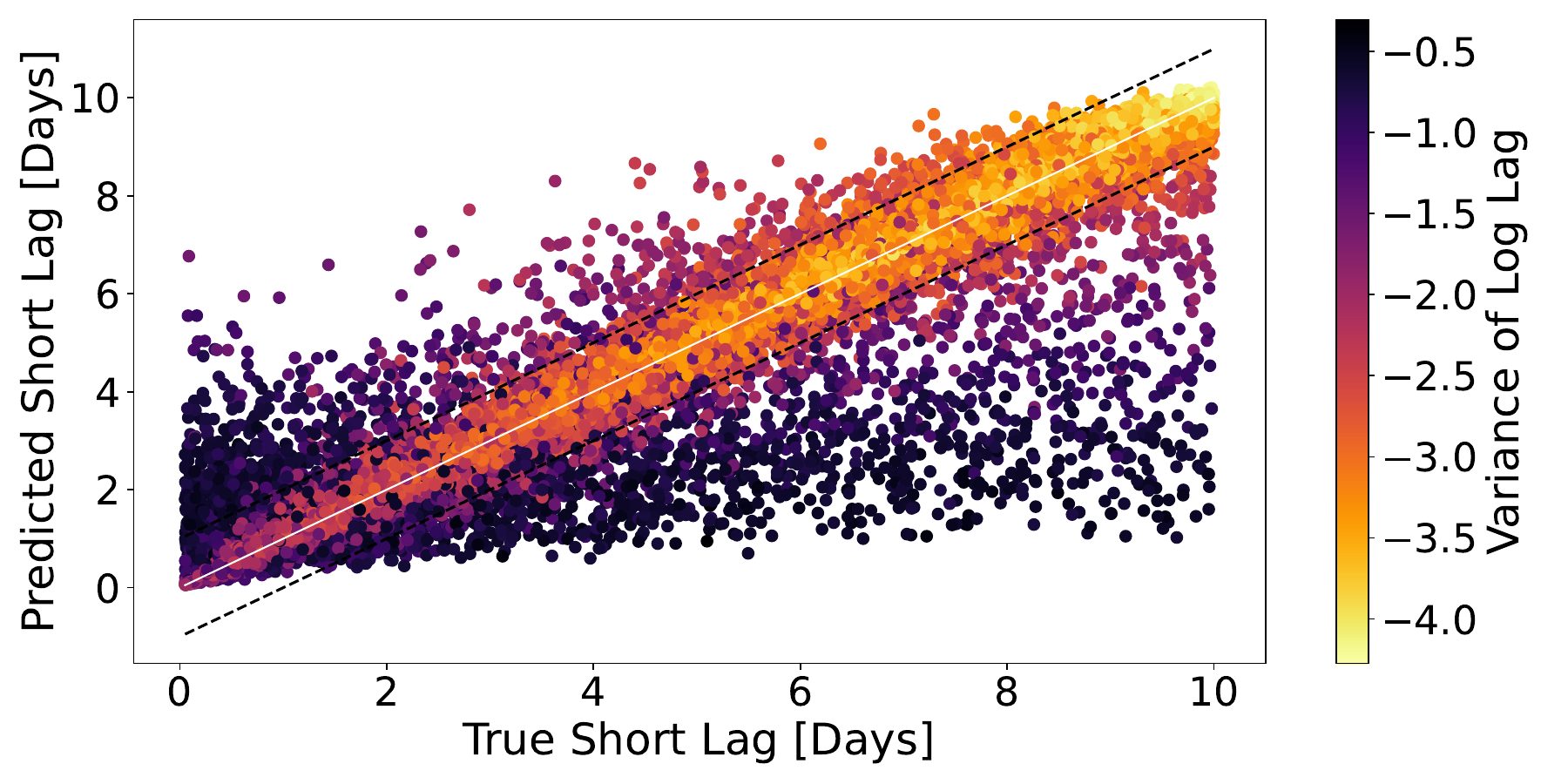} 
    \includegraphics[width=0.49\linewidth]{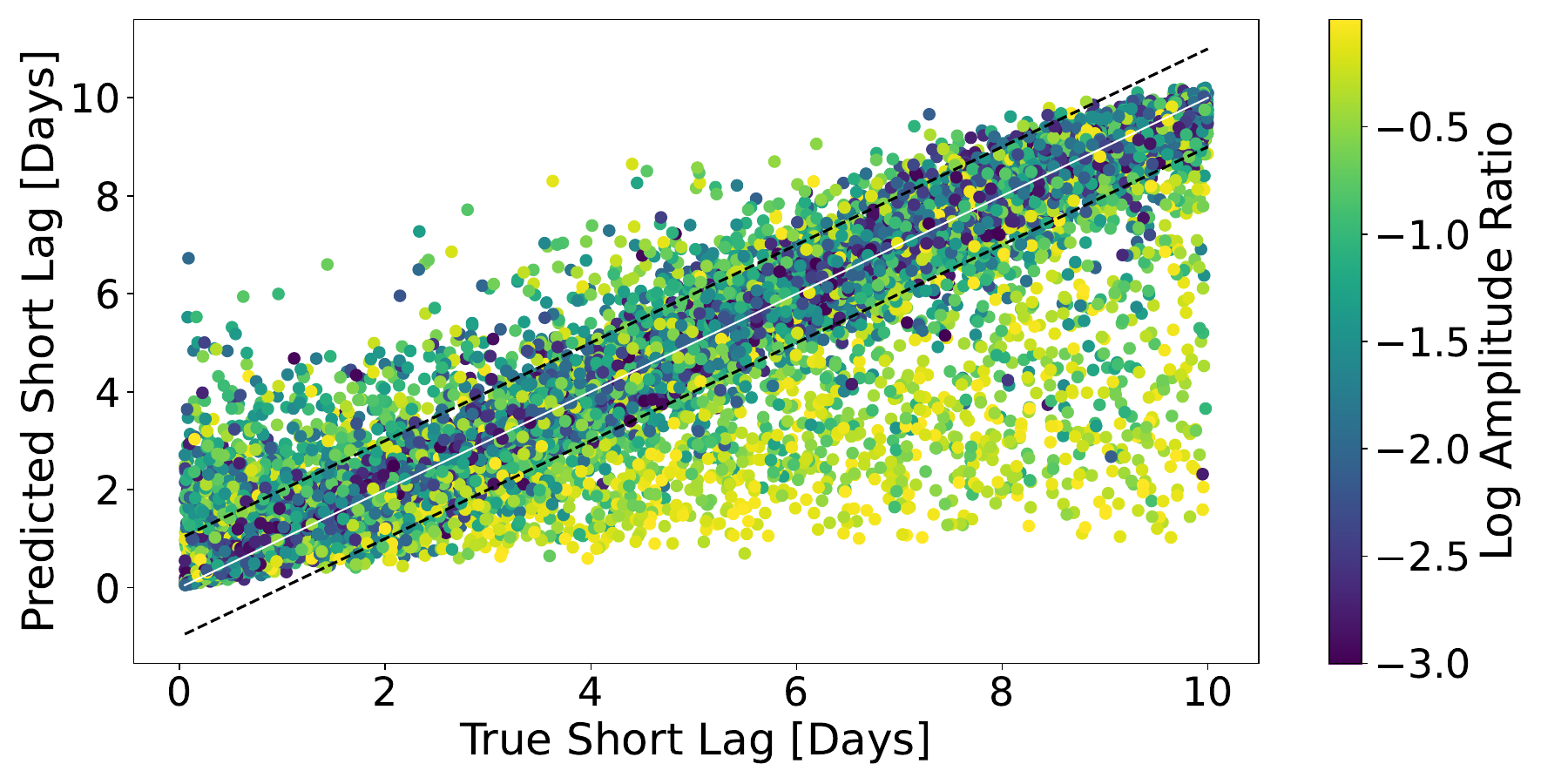} 
    \caption{The true short lag versus the short lag predicted by our transformer model for $2\times 10^4$ test set mock light curves. The solid white line is the one-to-one line and the dashed lines show $\pm 1$ day. In the left (right) panel points are color-coded by the variance of the posterior for the logarithm of the long lag (logarithm of the amplitude ratio). Our model is accurate to within 20\% of the true short lag for 79\% of light curves.}
    \label{fig:short}
\end{figure*}

\section{The Short Lag}
\label{app:short}
The main goal of our transformer model is to predict the presence and duration of long negative lags in AGN light curves. However, these same light curves will also have traditional disk continuum lags, or short lags, that our model can also predict. In this appendix, we study the ability of our model to accurately predict the short lag.

Figure \ref{fig:short} shows the true versus predicted short lags color-coded by the variance of the posterior for the logarithm of the short lag and the logarithm of the amplitude ratio in the left and right panel, respectively. Only 79\% of short lags predicted by our model are accurate to within 20\% of the true short lag, compared to 98\% of long lags predicted by our model. This inaccuracy is due to two main reasons. 

The first reason why the accuracy of our short lag predictions is lower than the accuracy of our long lag predictions is because we have combined observations down to a daily interval, and 10\% of our simulated short lags are less than 1~day long. We chose to take a daily average to help make the 10-year-long Rubin light curves more tractable and because intranight observations will not help us to detect long negative lags. If we exclude light curves with true short lags of less than 1~day, 83\% of short lags predicted by our model are accurate to within 20\% of the true short lag. Future work aimed at detecting short lags should refrain from averaging to a daily cadence for more accurate results.

The second reason for the lower accuracy is clear from the right panel of Figure \ref{fig:short}, which shows that the model performs badly on light curves with higher amplitude ratios. This poor performance makes sense, as the higher the amplitude ratio, the weaker the short lag signal. When we exclude light curves with $r>0.1$, 85\% of short lags predicted by our model are accurate to within 20\% of the true short lag. It is unclear if amplitude ratios $r>0.1$ exist in nature. If long lag signals this strong are common, we would more routinely detect them using the ICCF (see middle panel of Figure \ref{fig:main}) and the ICCF would fail to detect short lags more often than reported. For example, when we use the ICCF to find short lags in our test mock light curves with $r>0.1$ it is only accurate to within 20\% of the true short lag for 37\% of light curves.

If we eliminate both light curves with lags under 1~day and amplitude ratios $r<0.1$, we achieve 90\% accuracy. This accuracy could likely be improved by training our models for more iterations and on more light curves. \cite{Kovacevic:2022}, \cite{PozoNunez:2023}, and others found that LSST should be able to detect short lags in quasar light curves, and other machine learning codes work better for short lags \citep{Park:2021,Fagin:024, Li:2024}. Therefore, we refer to other existing methods for the detection of short lags, since the focus in this paper is on detecting long negative lags.

\section{Model Parameters}
\label{app:methods}

\begin{table*}[]
    \centering
\begingroup

    \caption{Summary of Mock Light Curve Parameters}
    \begin{tabular}{l|l|l}
        Parameter & Distribution & Range\\
        \hline
        $\tau_{\rm damp}$ [days] & $\mathcal{N}(\mu=107,\sigma=50)$ & $ [0.3,\infty)$\\
        SF$_{\infty}$ [mag] & Uniform & $  [0.01, 0.5]$\\ 
        $z$ & Uniform & $  (0, 4]$\\ 
        $\tau_{\rm long}$ [days] & Uniform & $  [-3650,-10]$, if $\tau_{\rm long}<-1825$, $\tau_{\rm long}=0$\\
        $\tau_{\rm short}$ [days] & Uniform & $  [0.05, 10]$\\
        $S_{\rm long,short}$ [days] & $\mathcal{N}(\mu=\tau_{\rm long,short}/5,\sigma=\tau_{\rm long,short}/5)$ & $ [\tau_{\rm long,short}/10,\tau_{\rm long,short}/2]$\\
        $r$ & Log-Uniform & $[10^{-3},1]$ \\
        $m$ & $\mathcal{N}(\mu=20.9,\sigma=1.04)$ & $(-\infty,\infty)$
    \end{tabular}
    \label{tab:params}
    \endgroup
    \\
    \bigskip 
    \textbf{Notes.} For each mock light curve parameter in the left column, we provide the random distribution in the center column (normal, uniform, or log-normal) and the distribution range in the right column.
\end{table*}

\begin{table}[]
\centering
    \begingroup
    
    \caption{Model Hyper-Parameters}
    \begin{tabular}{l|l|l}
        Hyper-Parameters & Range & Optimal\\
        \hline
        heads & 2, 4, 8, 16 & 16\\
        layers & 2, 4, 6, 8, 12 & 8\\
        dimensions & 64, 128, 256, 512, 1024 & 256\\
        learning rate & $10^{-4}-10^{-2}$ & $10^{-3}$\\
        warm-up steps & $100-5000$ & 3,112 \\
        warm-up rate multiplier & $0.01-1$ & 0.7 \\
        training steps & $10^{4}-10^{5}$ & 98,413 \\
    \end{tabular}
    \label{tab:hyper}
    \endgroup
    \\
    \bigskip 
    \textbf{Notes.} The hyper-parameter grid for our 75 wandb trials (center column) and the optimal parameter (right column).
\end{table}

The parameter distributions we draw from to generate $\num{2e6}$ training, $\num{2e4}$ validation, and $\num{2e4}$ test mock Rubin AGN light curves are given in Table \ref{tab:params}. For each DRW driving light curve, we randomly draw the damping timescale, $\tau_{\rm damp}$, and the structure function at infinity, SF$_{\infty}$, from a normal and uniform distribution, respectively. We generate our DRW driving light curves with a high cadence of $dt=0.01$~days and a 200~year baseline. We then divide these driving light curves into 10 light curves. We take this approach because it is important when generating a DRW light curve that the baseline is significantly longer than the damping timescale \citep{Kozlowski:2017, Stone:2022}.  

In addition to the lag parameters mentioned in Section \ref{sec:methods:sims}, the width, $S$, of the long and short lag Gaussians are drawn from a random normal distribution centered at 20\% of the mean lag. We also draw a redshift, $z$, and magnitude, $m$, from a random uniform and normal distribution, respectively. $z \in (0,4]$ and we choose $\mu=20.9$ and $\sigma=1.04$ for the mean and standard deviation of the magnitude distribution to span typical apparent magnitudes for AGN we expect to observe with Rubin. 

To optimize our transformer model, we perform a Bayesian optimization for the hyper-parameter ranges given in the center column of Table \ref{tab:hyper} with 75 trials on wandb using the Tree-structured Parzen Estimator algorithm to pick the next point \citep{Bergstra:2011}. We give the value of each parameter for our model that minimizes the loss in a validation set of $\num{2e4}$ mock light curves in the right column of Table \ref{tab:hyper}.

\bibliography{agn_lag.bib}
\end{CJK*}
\end{document}